# Effect of Proxy Nodes on the Performance of TCP-Based Transport Layer Protocols in Wireless Sensor Networks

**Fatemeh Sadat Tabei[1] and Behnam Askarian[2]**

[1, 2] Department of Electrical and Computer Engineering, Texas Tech University, Lubbock, TX 79409, USA

[1]fatemehsadat.tabei@ttu.edu, [2]behnam.askarian@ttu.edu

**ABSTRACT**

Wireless Sensor Networks (WSNs) have recently attracted many researchers' attentions due to their wide range of applications. Even though a plethora of studies have been carried out on characteristics, special conditions, and various aspects of WSNs, t. ransport protocol which is compatible with conditions of WSNs has not been considerably addressed. WSNs have limitations such as storage space, energy resources, and wireless communication issues. Accordingly, widely-used transport protocols like Transmission Control Protocol (TCP) may not enjoy sufficient efficiency in such networks. In this paper, we study the characteristics of WSNs leading to design transport layer protocol for WSNs and aim at evaluating the efficiency of TCP and its dependent protocols (TCP variables), which are introduced to wireless networks. We propose to employ proxy nodes near sinks to improve the performance of transport layer. Our NS-2 simulation results indicate that throughput and packet delivery ratio are improved from 20 to 50 percent after employing proxy nodes, while the average message delay is almost increased twice.

## 1. INTRODUCTION

Wireless Sensor Networks (WSNs) are a series of sensor nodes which seek to cooperatively gather environmental data and transfer them to a center (main station). Nowadays, they are found to have various applications in numerous fields [1]. The wireless sensor nodes can assess physical phenomena, process the sensed data locally, and pass it to a central station in a raw or collected form. There have been studies related to data link and network layers in WSNs [2-10]. However, few studies have been carried out on transport layer even though the guarantee of receiving data from nodes to central stations is of high significance under limitations of WSNs due to lack of energy power, storage space, and noisy wireless communications. Studies conducted on transport layer of wireless networks mostly focus on two issues, namely, reliable data delivery and congestion control. Quantitative protocols such as RCRT [11] and STCP [12] were proposed to cope with these issues.

Protocols focusing on reliable data delivery can be categorized in terms of the direction of data transfer. Node-to-sink reliability protocols include RMST [13], RBC [14], DTSN [15], and FLUSH [16] while sink-to-node reliability protocols are PSFQ [17], PALER [18], GARUD [19], and HRS [20]. These proposed protocols usually either improve delivery reliability or enhance congestion control issues. However, despite taking necessary measures and actions, a general transport layer protocol, which enjoys high levels of efficiency in WSNs for event-oriented applications or other ones, have not been paid enough attention.

Another problem related to the introduced protocols is testing the proposed method in different and uneven applications and test-bed and using special motes. As a result, the evaluation of previously conducted studies cannot be compared with each other [21]. TCP is one of the most popular and applicable transport layer protocols in normal networks. Due to its complexities, this protocol cannot be employed in sensors with limited wireless communication capacity [12]. Accordingly, to resolve this issue, variables of TCP are introduced, the most popular of which are Reno, NewReno and Vegas.

In this paper, we propose to deploy proxy nodes to improve the performance of transport layer protocol in WSNs. The proposed method passes the data to main station node using TCP protocol and TCP variants. The present study examines the degree of performance in two different states: 1) proxy nodes as an intermediary node, and 2) proxy nodes are neighbor of sink. Our results show that deploying proxy nodes near sinks helps the network identify packet loss or congestion before the data is delivered to sink.

The rest paper is organized as follows. In Section II, we address the main criteria for designing transmission protocol. TCP variants for wireless networks are explained in section 1. Section 2 explains the network design with proxy nodes and, section 3 presents results and analysis. Finally, Section 4 concludes this paper.

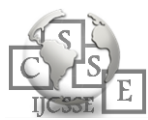

## 2. WSN TRANSPORT LAYER DESIGN CONSIDERATIONS

Transport layer protocols are responsible for prioritization, data segmentation, transition data flow control, congestion control, recovery of lost packets, and guaranteeing quality of service. Contrary to traditional TCP/IP networks, each node in WSNs has a limited energy, bandwidth, and storage space and needs to be able to cope with noisy wireless channel.

A reliable transport protocol is required to be robust and compatible with various scenarios like node failures and route changes. The process of protocol should be simple and fast [22]. The criteria used for evaluating the performance of WSNs transport protocol are as follows:

- *Reliability*: In WSNs, reliability can be evaluated in both packet level and event level. Packet level reliability is measure by the ratio of packets which are successfully delivered to the final destination. The alternative way of calculating packet level reliability is to compute end-to-end loss rate of sum of packets. The less the loss rate, the better the network reliability is. Event level reliability is measured by counting a certain amount of data from target or event by the receiver. In this state, all packets need not to be received by receiver. Rather, as long as a certain percent of packets reaches the destination in a certain period of time, event level reliability is achieved.
- *Quality of Service*: The quality of service encompasses parameters such as used bandwidth, network latency, on time and in order delivery, and operational power. Quality of service requirements vary correspondingly to different applications of WSNs.
- *Energy Efficiency:* The required energy for WSNs is supplied through a battery. Tee WSNs may expand in non-urban areas where the possibility of data collection with low energy consumption is regarded as an important efficiency index. Moreover, energy efficiency can be calculated by computing the overall consumed energy in the network under which the required percentage of reliability is maintained.
- *Reliable data delivery*: Reliable data delivery in WSNs is a vital issue and may vary from one application to another. There can be different design choices for data transport protocol in WSNs. For instance, the protocol can be on the basis of end-to-end or hub-to-hub structures. Also, the structure of lost data recovery is based on positive or negative delivery and reliable data delivery can be introduced in a node-to-sink or sink-to-node direction.
- *Congestion control*: Congestions can occur in WSNs for various reasons: simultaneous data transport, addition, removal of sensor node, and consecutive messages resulted from various events [23], [24].

Network congestion can bring about two serious results: buffer space drop and increase in the cost of resources for each packet. Hence, decreasing congestion is beneficial to achieving reliability. In network sensors with one sink node, decrease in congestion can be achieved by implementing passive method. Also, rate control is a widely-used method [25]. When congestion is identified in a system, sensor nodes decrease their reporting process, providing the opportunity for congested nodes to leave their line and become released.

## 3. TCP VARIANTS FOR WIRELESS NETWORKS

Researchers have proposed some variables for TCP to resolve some of the mentioned issues in the last section. Here we review three popular TCP variants including Reno, NewReno and Vegas.

- TCP Reno Protocol: Reno protocol supports fast-recovery, header prediction options, and delayed delivery. Its main advantage is that it keeps new data entry time by twofold deliveries and also prevents from entry into slow starting phase when TCP transfer rate is decreased. The degree of improvement in this method is significant, particularly in connections where multiplication of bandwidth in delay is high since slow starting phase takes much longer time in this state. In high-speed links, any kind of normal congestion can cause several parts to be lost. In this case, re-transfer and fast-recovery wouldn't be capable of retrieving lost parts. Thus, slow starting mechanism is repetitively recalled.
- TCP NewReno Protocol: NewReno presents a new fast-recovery phase so that whenever fast re-transfer starts for the first time, sender of the last number keeps the recovery order. This protocol can manage packets loss in only one fast-recovery phase whereas Reno needs to recall fast-recovery multiple times.
- TCP Vegas Protocol [26]: TCP Vegas which is invented based on Reno improvements is a complement to TCP. It employs three methods (New Retransmission Mechanism, Congestion Avoidance Mechanism, and Modified Slow-Start Mechanism) to improve operating power and decrease the number of lost packets. By comparing calculated and expected operational powers in each window, Vegas protocol monitors changes in operational power.

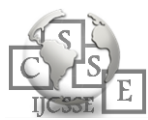

In this paper, the performance of TCP and three mentioned TCP variants for wireless networks are investigated in WSN proxy based network.

## 4. TRANSPORT LAYER PROTOCOLS USING PROXY NODES

The goal of this paper is to study the performance of TCP and its named variants in WSNs. Moreover, by dividing the network and making changes in sending data to the central station, variations in average network overall throughput, average end-to-end delay, and average packet delivery ratio are studied.

In our proposed method, the network is divided into sections. In each section, one proxy node plays the role of the central station of that section while the other nodes send the sensed data to this node instead of sending it to sink directly. Proxy node is an intermediary node that acts as both sink and also simple node for the purpose of sending sensed data on behalf of other nodes. Proxy node is responsible for receiving data from other nodes in the section and passing them to the central station of the entire network. Previous studies mainly focused on introducing protocols to WSNs or varying applications of WSNs.

At first, a network with above-mentioned characteristics and also with moving nodes (except for the central station) is designed as the basis of comparison with the proposed design. Then, the network performance status including success in passing packets, end-to-end delays, and the ratio of packet reception for four transport protocols like TCP and its variants (Reno, NewReno, and Vegas) are examined. In the proposed deploying proxy nodes, end-to-end sending method is changed and the main network is divided into separated areas, in each of which a node is considered as the central node receiving packets of that area. At first, packets of each area begin to send their packets to the central node of their area (proxy node). Then, after collecting all packets, proxy node passes them to the central station.

In this method, collected data is passed to destination and it may cause packets reception time to the main station increases. By the way, this will help receiving packet loss to decrease.

One important parameter in designing the network is a number of nodes. The selection of this parameter should be considered not to make complex conditions of wireless networks more complicated.

The number of sensor nodes in a network is determined based on the network performance and area under employment. For instance, in house applications, covered area and the number of nodes are highly restricted while in external use such as agricultural applications, the covered area is wider and the number of nodes is larger. In most of the previous studies, the number of nodes ranges between 100 and higher numbers [21].

In this examination, nodes traffic in three different states is taken into consideration. Also, the proposed system is analyzed in two methods: in the first one, with regard to density, selected proxy node in each area should be in the middle of that area. In the second one, with regard to density, selected proxy node should be one of the nodes close to the main station.

## 5. RESULTS

We evaluated the effect of proxy existence on the performance of TCP protocols in WSNs for varying the network size and the mobility of nodes. The effect location of proxy on the performance of TCP protocols is also investigated. As performance metrics, we considered network overall throughput, end-to-end delay, and packet delivery ratio defined as follows:

- Average network overall throughput: the average number of bits which successfully reach their destination per every source-destination pair called throughput. The sum of these throughputs is the network's overall throughput [27].

- Average end-to-end delay: time spent by a packet to reach destination is its delay. The delay depends on several factors in the network such as the number of nodes, nodes' transmit power, and the network traffic structure [28], [29].

- Average packet delivery ratio: the ratio of packets received by destinations to total number of packet sent by the sources is the network called packet delivery ratio. The larger this ratio, the better it shows the network and protocol efficiency.

The selection of the number of proxy nodes should be determined in such a way that transport protocol be able to control the packets sent in the network. Here, we evaluated the performance with the network size of 50, 100 and 110 nodes.

In addition to the number of nodes, network scale is another factor affecting design. In the present study, a network with a length of 1000 m and width of 1000 m is taken into account. The transmission radio range is 100m. This network is a wireless one, in which nodes' location is assumed as being uniform random distribution. The goal of the network nodes is to pass their packets to the central station.

*Table 1: Ns-2 Simulation PaRAMETERS*

| Area of sensor field | 1000*1000 m² |
|---|---|
| Number of nodes | 50 |
| | 100 |
| | 110 |
| Number of proxy nodes | 4 |
| Radio range of a sensor node | 100m |
| Routing Protocol | AODV |
| MAC Protocol | IEEE 802.11 |

Nodes are assumed moving and their movement and velocity characteristics are assumed as random. Thus, the network's overall status can't be predicted during time period. Protocols IEEE 802.11[30] and AODV [31] are used for MAC layer and routing protocol, respectively. Moreover, built in patches of TCP variants that are added in NS-2.34 are used in this study.

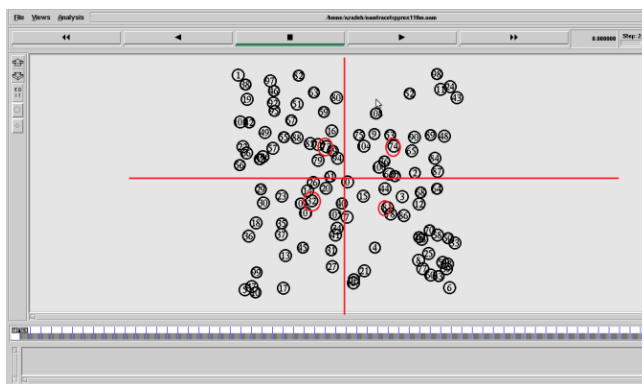

*Fig. 1. Proxy-Based network with 110 nodes and proxy nodes are neighbor of sink.*

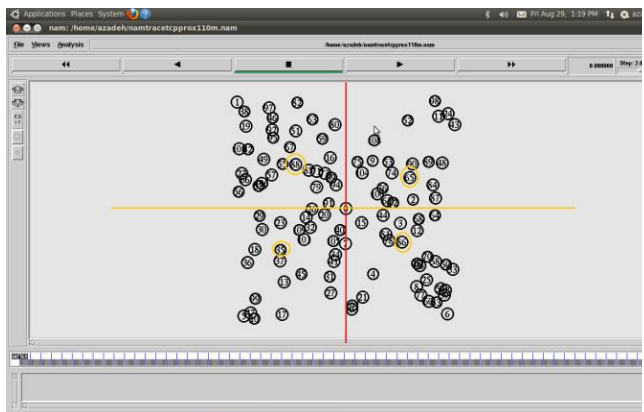

*Fig. 2. Proxy-Based network with 110 nodes and proxy nodes are in the middle.*

## 5.1 Non-Proxy Network Simulation Results

Here we study the performance of transport protocols without proxy nodes. The simulations results are demonstrated in Table I.

The results show that as the number of nodes in the network and traffic increases, the throughput of TCP, Reno, and NewReno protocols in the non-proxy system decreases. Also, the performance these three protocols are only slightly different. Vegas protocol shows a poorer performance compare to other transport protocols. Particularly, the throughput of Vegas is significantly lower than that of other ones.

Having an active behavior, Vegas protocol prevents packets from being lost in the network. Due to such nature, it limits the amount of data it can transfer in the network. Contrarily, Reno and NewReno protocols re-send packets which are repetitively lost. Therefore, they make Vegas protocol show a poorer performance in terms of the network throughput.

The end-to-end delay difference for all four protocols is not significant. Moreover, it is observed that end-to-end delay for Vegas protocol is lower than that of other protocols due to its fast re-send method. Vegas protocol makes use of an effective algorithm to examine the delay in program. Also, it has overcome the issue relating to enough amount of positive reception to identify lost packets. Therefore, with regard to end-to-end delay factor, which is of high importance in WSN applications, Vegas protocol performs better than other ones.

With increase in the number of network nodes, TCP performance is poorer compared to other protocols, dramatically dropping. The packet delivery ratio of TCP is 0.9703 while it is 0.9729 for Reno, and 0.9741 and 0.9763 for New reno and Vegas respectively for 110 nodes in the network. Moreover, the packet delivery ratio of TCP drops from 0.9756 with 50 nodes in the network to 0.9703 for 110 nodes in the network. This issue alongside with this fact that TCP does not have a better performance in terms of other evaluation criteria shows that TCP is inefficient for WSNs, particularly when density of nodes in network grows.

*Table 2:Network without proxy nodes*

| Transport protocol | Number of nodes | Throughput (Kbps) | End-to-End Delay (ms) | Packet Delivery Ratio |
|---|---|---|---|---|
| TCP | 50 | 177.84 | 120.599 | 0.9756 |
| | 100 | 311.29 | 163.531 | 0.9721 |
| | 110 | 301.17 | 137.744 | 0.9703 |
| Reno | 50 | 185.96 | 111.832 | 0.9807 |
| | 100 | 321.27 | 146.119 | 0.9744 |
| | 110 | 136.28 | 136.641 | 0.9729 |

| Transport protocol | Number of nodes | Throughput (Kbps) | End-to-End Delay (ms) | Packet Delivery Ratio |
|---|---|---|---|---|
| NewReno | 50 | 180.93 | 118.187 | 0.9747 |
|  | 100 | 309.41 | 165.13 | 0.9718 |
|  | 110 | 351.41 | 179.893 | 0.9741 |
| Vegas | 50 | 80.09 | 99.388 | 0.9805 |
|  | 100 | 151.27 | 92.753 | 0.9743 |
|  | 110 | 179.89 | 174.82 | 0.9763 |

### 5.2 Proxy-Base Network Simulation Resulst

Now we study the performance of transport protocols when some proxy nodes are deployed in the network.

Comparison of the results of network throughput in two proxy and non-proxy states shows that the network throughput experiences a considerable increase in TCP, Reno, and NewReno protocols. However, Vegas protocol has still poorer performance. Meanwhile, when the proxy nodes are deployed in the network, the throughput of Vegas protocol shows a slight improvement.

In proxy state (middle and neighboring to sink nodes), the performance of NewReno protocol in terms of the network throughput is more favorable in comparison with that of other protocols. However, adding proxy to the network data transmission process improved its throughput, it also increased end-to-end delay. Comparing packet delivery ratio in the proxy mode show that this criterion has improved in Vegas protocol. In the contrary, the same criterion for other protocols experienced a drop and the proxy system with middle nodes is in the worst status. Vegas protocol does not show a favorable performance in terms of the network throughput with other protocols due to its nature. However, with regard to the network throughput and end-to-end delay in each three systems, it has a favorable performance. On the other hand, by adding proxy to other protocols, a dramatic increase in end-to-end delay was resulted. This increase is because data is first transmitted to the proxy node, then it is sent to the sink. This adds up the time of the transmission, but improves other performances. End-to-end delay was a lot lower in Vegas protocol than other protocols due to its fast re-send algorithm.

*Table 3: Network with proxy: proxy nodes are in the middle.*

| Transport protocol | Number of nodes | Throughput (Kbps) | End-to-End Delay (ms) | Packet Delivery Ratio |
|---|---|---|---|---|
| TCP | 50 | 280.43 | 236.644 | 0.9389 |
|  | 100 | 364.88 | 569.974 | 0.9641 |
|  | 110 | 431.46 | 414.309 | 0.9653 |
| Reno | 50 | 280.64 | 262.425 | 0.9455 |
|  | 100 | 379.06 | 405.946 | 0.9642 |
|  | 110 | 426.13 | 471.149 | 0.9660 |
| NewReno | 50 | 241.807 | 250.80 | 0.9451 |
|  | 100 | 370.73 | 580.75 | 0.9666 |
|  | 110 | 425.29 | 468.011 | 0.9659 |
| Vegas | 50 | 130.20 | 164.993 | 0.9758 |
|  | 100 | 184.55 | 327.415 | 0.9766 |
|  | 110 | 204.99 | 222.381 | 0.9729 |

*Table 4: Network with prox: proxy nodes are neighbor of sink.*

| Transport protocol | Number of nodes | Throughput (Kbps) | End-to-End Delay (ms) | Packet Delivery Ratio |
|---|---|---|---|---|
| TCP | 50 | 271.41 | 407.763 | 0.9483 |
|  | 100 | 367.49 | 649.238 | 0.9649 |
|  | 110 | 441.99 | 800.84 | 0.9674 |
| Reno | 50 | 256.56 | 576.858 | 0.9541 |
|  | 100 | 384.59 | 447.335 | 0.9650 |
|  | 110 | 446.81 | 835.774 | 0.9724 |
| NewReno | 50 | 269.33 | 468.95 | 0.9490 |
|  | 100 | 382.66 | 527.59 | 0.9638 |
|  | 110 | 444.08 | 748.50 | 0.9685 |
| Vegas | 50 | 135.67 | 210.925 | 0.9805 |
|  | 100 | 185.75 | 30.6.03 | 0.9753 |
|  | 110 | 213.40 | 281.852 | 0.9722 |

## 6. CONCLUSION

The present study aimed to examine and introduce a method of improving transport protocol in WSNs. Firstly, it addressed the performance of TCP is poor, compared to that of Reno, NewReno, and Vegas protocols. Also, according to simulation results, though Vegas protocol performs poorly in terms of network throughput, it shows a much better performance than TCP, Reno, and NewReno with regard to end-to-end delay and packet delivery ratio.

Given limitations in WSNs and also in introducing an appropriate transport protocol, the use should be made of simple and general methods applicable to all WSNs. Proxy method was an idea presented in this study. If we tend to use end-to-end sending and receiving method, the odds are that as nodes traffic in the network increases, reliability of packet delivery to the main station decreases. But, in proxy state, when packets were sent to the proxy node, in case a packet was congested or lost, it was rapidly identified and packet recovery or congestion prevention operation was employed on the basis of transport protocol structure (Reno, NewReno, and Vegas).

In other previously examined methods, simulation environment was selected on the basis of certain motes. Thus, the introduced protocol had a special dependency on mote structure or certain application of WSN. But, in this state, the network overall performance and efficiency was examined while methods proposed in the past only addressed the issue of improving a certain parameter such as prevention of congestion or recovery of lost packets.

Our future work concerns about the effect of increasing the number of proxy on the performance of the network, also which parameters should we concern about selecting proxy nodes. Following completion by checking other parameters like changes in congestion window parameter (CWND) can also be a part of our future work.

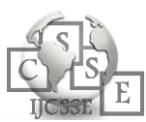